\documentclass[article,a4paper,showpacs]{revtex4}
\usepackage{graphicx}

\begin{document}

\title{On resumming periodic orbits in the spectra of
       integrable systems}

\author{Alfredo M.~Ozorio de Almeida\dag\ddag, Caio H.~Lewenkopf\S\ 
        and Steven Tomsovic \dag$\|$}

\address{\dag\ Max-Planck-Institut f\"ur Physik komplexer Systeme,
               N\"othnitzer Str.~38, 01187 Dresden, Germany}
\address{\ddag\ Centro Brasileiro de Pesquisas F\'{\i}sicas, R. Xavier
                Sigaud 150, 22290-180 Rio de Janeiro, Brazil}
\address{\S\ Instituto de F\'{\i}sica, Universidade do Estado do Rio de
             Janeiro, R. S\~ao Francisco Xavier 524, 
             20559-900 Rio de Janeiro, Brazil}
\address{$\|$\ Department of Physics, Washington State University,
             Pullman, WA99164-2814, USA}

\date{\today}

\begin{abstract}
Spectral determinants have proven to be valuable tools for resumming the
periodic orbits in the Gutzwiller trace formula of chaotic systems.  We
investigate these tools in the context of integrable systems to which
these techniques have not been previously applied. Our specific model is a
stroboscopic map of an integrable Hamiltonian system with quadratic action
dependence, for which each stage of the semiclassical approximation can be 
controlled. It is found that large errors occur in the semiclassical traces 
due to edge corrections which may be neglected if the eigenvalues are obtained 
by Fourier transformation over the long time dynamics.  However, these errors
cause serious harm to the spectral approximations of an integrable system
obtained via the spectral determinants.  The symmetry property of the spectral 
determinant does not generally alleviate the error, since it sometimes
sheds a pair of eigenvalues from the unit circle.
By taking into account the leading order asymptotics of the edge corrections, 
the spectral determinant method makes a significant recovery.
\end{abstract}

%  03.65.Sq  ... semiclassical theories and applications
%  02.30.Ik  ... integrable systems
%  02.30.Lt  ... sequences, series and summability

\maketitle

%%%%%%%%%%%%%%%%%%%%%%%%%%%%%%%%%%%%%%%%%%%%%%%%%%%%%%%%%%%%%%%%%%%%%%%%%%%%%%%%%%%%%
\section{Introduction}
\label{intro}

It has been generally accepted that spectral determinants, or
zeta-functions, provide optimal semiclassical estimates of the individual
energy levels of classically chaotic systems~\cite{cv,Cvitanovic89,Eckhardt89}.  
These resummations of the periodic orbits in the Gutzwiller trace
formula~\cite{Gutzwiller71,Balian72}, originally developed for time-independent
systems, may be obtained in a variety of ways.  The formulation of Berry and 
Keating~\cite{Berry90JPA} relies explicitly on the instability of the periodic 
orbits in such a way as to give an expression that does not depend on a sharp 
cut-off of the orbit periods. 
In contrast, Bogomolny's approach~\cite{Bogomolny90,Bogomolny92} reduces the 
problem to the quantization of a map over a Poincare\'e surface of section, 
without making any assumption about the nature of the classical motion. However, 
two sharp boundaries are introduced. On the one hand, the Poincare\'e section 
itself is bounded if the constant energy surface is compact. The limited area 
of the section, which corresponds to a finite dimension of the Hilbert space, 
then leads to a sharp cut-off in the period of the orbits.
The recent paper of Eckhart and Smilansky~\cite{Eckhardt01} also works with a 
quantum map in a bounded region, but this is obtained stroboscopically instead 
of with a Poincar\'e section.  

The classical motion of integrable systems is restricted to invariant tori, which
determine closed invariant curves (also tori) of the Poincar\'e mapping.
For integrable systems, Berry and Tabor~\cite{bt} established the
equivalence of the Gutzwiller trace formula to the general forms of
Bohr-Sommerfeld quantization.  The latter method is evidently the most
efficient for the calculation of individual levels.
Nevertheless  the exercise of showing the equivalence of both methods
had the merit of clarifying the role of periodic orbits (forming continuous
tori for higher dimensional systems) in the density of states.  
Indeed it raises an important question: 
The Berry-Tabor equivalence involves the complete set of periodic orbits,
so how might one obtain correct energy levels from the resummation of a
finite selection of short orbits?  

Our interest in this paper is resummation approximations for the spectra
of simple integrable systems.  Particularly, we will focus on the effect 
of introducing boundaries that do not interfere with the local classical 
tori. For simplicity, consider such a system with a single degree of freedom.  
For the classical Hamiltonian $H(I)$, where $(I,\theta)$ are the action-angle 
variables~\cite{text}, the Bohr-Sommerfeld levels are
\begin{equation}
\label{bs}
E_n=H\left(\hbar\left(n+{1 \over 2}\right)\right) \ .
\end{equation}
Actually, it suffices to consider a case for which the action variables
are those of the harmonic oscillator
\begin{equation}
\label{ho}
I={1\over 2}(p^2+q^2) \ ,
\end{equation}
rendering the semiclassical approximation of \ (\ref{bs}) exact.  It
is important to point out that large-$n$ states are well approximated by
\ (\ref{bs}) even for a general nonquadratic dependence of the
Hamiltonian on the phase space variables.  The main question
is how well a spectral determinant method converges to the
eigenvalues $E_n$ in this case.  

A conceptually clean approach to answering this question follows in the
spirit of Creagh~\cite{Creagh95}, who applied an extension of resummation
methods~\cite{Saracero92,Smilansky93JPA} to the perturbed cat maps.  He found 
worsening errors
in the semiclassical trace formula as the perturbation brought the map
out of the hyperbolic regime, but was unable to continue the perturbation
all the way to an integrable regime.  Here, since $I$ is a constant of the
classical evolution, our consideration can be limited to a ring
${\cal I}_-<I<{\cal I}_+$, thus obtaining a finite phase space and hence a
finite Hilbert space of dimension $N=\{({\cal I}_+-{\cal I}_-)/\hbar \}$,
where $\{\ \cdot \ \}$ denotes the integer part.  A stroboscopic map of
period $\tau$ can be defined and quantized in such a way that the dynamics
is determined by the evolution operator 
\begin{equation}
\label{un}
U_N = \sum_{n=N_-}^{N_+} \exp\!\left(-{i\over \hbar}E_n\tau\right)
|n\rangle\langle n |  \ .
\end{equation}
The quantity $\hbar N_-$ ($\hbar N_+$) is the smallest (largest) quantized action 
greater (smaller) than ${\cal I}_-$ (${\cal I}_+$) and $|n\rangle$ are the eigenstates.  
The definition of the
spectral determinant as $\det\left(1-zU_N\right)$ gives an $N^{th}$ order
polynomial in the variable $z$ and the roots of the characteristic
equation
\begin{equation}
\label{char}
P_N(z)\equiv \det\left(1-zU_N\right) =0
\end{equation}
are just
\begin{equation}
\label{ev}
z_n=\exp\left({i\over \hbar}E_n\tau\right) \ .
\end{equation}
Thus, knowledge of the phases $\phi_n$ of the zeros of the spectral
determinant, or zeta-function, determines the eigenvalues of the
stroboscopic map within the ring. These coincide with a subset of the
eigenvalues of the original system possessing an infinite Hilbert space.

The eigenenergies of the original system can be inferred from
\begin{equation}
\label{infer}
E_n={\hbar\over \tau}(\phi_n +2k\pi) \ .
\end{equation}
The determination is unique if, for the energy interval of the ring
$(E_-,E_+) = (E({\cal I}_-),E({\cal I}_+))$, the constraint  
\begin{equation}
\label{constraint}
\left(E_+-E_-\right){\tau\over \hbar} \le 2\pi
\end{equation}
is respected.  This scheme for locating eigenvalues of the original
Hamiltonian resembles that of~\cite{Eckhardt01}, but our objective of comparing
semiclassical methods can be achieved also in the context of the
continuous dynamical system confined to the ring, or even the discrete
map defined by (\ref{un}).

In the semiclassical limit, $\mbox{Tr}(U_N^l)$ is expressed as a sum over
periodic orbits of period $l$ for the classical stroboscopic map
generated by the Hamiltonian $H(I)$ in the time $\tau$ within the ring. 
These orbits, with action $I_{m,l}$, such that
\begin{equation}
\label{deriv}
\tau {{\rm d}H(I)\over {\rm d}I} = {2\pi m\over l} \ ,
\end{equation}
define continuous curves, unless $I=0$.  To simplify the treatment, we
consider ${\cal I}_->0$, which excludes the isolated periodic orbit at the
origin.  

Note that the periodic orbits of the stroboscopic map comprise only a
small subset of those of the original system with continuous time. 
Indeed, all the orbits of the latter are periodic whereas only those full
orbits with periods that are rationally related to the stroboscopic time,
$\tau$, are periodic in the map.  The map resembles a Poincar\'e
map of an integrable system with two degrees of freedom in the sense that
all orbits lie on invariant curves, but the curves made up of periodic
orbits form a dense set of zero measure.

The semiclassical form for the spectral determinant is obtained from
the expression 
\begin{equation}
\label{specdet}
\det\left(1-zU_N\right)=\exp \left[ - \sum_{l=1}^\infty {z^l\over l}
\ \mbox{Tr}(U_N^l) \right] \ .
\end{equation}
Even though the series only converges for $|z|<1$, it is possible to
follow~\cite{Saracero92} in noting that the Taylor expansion of the
exponential in Eq.\ (\ref{specdet}) can be identified with the finite
expansion of the spectral determinant (\ref{char}):
\begin{equation}
\label{expan}
P_N(z)= 1 + c_1 z + c_2 z^2 ... + c_N z^N \ ,
\end{equation}
where the coefficients are given by the following recurrence relation
\begin{equation}
\label{eq:recurrence}
c_k = - \frac{1}{k} \sum_{l=1}^k c_{k-l} \mbox{Tr}\,(U_N^l) \ .
\end{equation}
In this way the periodic orbits up to period $l=N$ determine, in
principle, all the eigenvalues of $U_N$.

A further reduction of the period of the orbits used in the semiclassical
spectral determinant results from the symmetry relation for the
coefficients~\cite{Bogomolny90} 
\begin{equation}
\label{symm}
c_k=(-1)^N\det(U_N) c_{N-k} \ .
\end{equation}
The coefficients containing traces for long iteration times, $l>N/2$, can be obtained from
the coefficients with $l<N/2$. Hence, only the orbits of period $l<N/2$ are needed. 
This is a particularly simple example of ``bootstrapping''~\cite{Berry85}
resulting from the finiteness of the Hilbert space.  For chaotic systems,
the symmetry (\ref{symm}) is a fundamental tool because the number
of periodic orbits increases exponentially with $l$.  This is not so for
integrable systems.  

In Section~\ref{st}, we derive the semiclassical formula for $\mbox{Tr}(U^l)$ of
the integrable map defined in \ (\ref{un}). It will then become clear that 
edge corrections to the periodic orbit sum cannot be neglected in this case.  
Indeed these are the only corrections for the particular model that we study
numerically in Section~\ref{model}. 
Though the absence of edge corrections severely affects the spectral determinant, 
which has a natural cut-off in time, they cancel out in the Berry-Tabor formula 
as discussed in Section~\ref{equiv}.

%%%%%%%%%%%%%%%%%%%%%%%%%%%%%%%%%%%%%%%%%%%%%%%%%%%%%%%%%%%%%%%%%%%%%%%%%%%%%%%%%%%%%
\section{The semiclassical trace}
\label{st}

Relation~(\ref{bs}) for the eigenvalues becomes exact in the case of the
harmonic oscillator actions (\ref{ho}). Likewise, the propagator (\ref{un}), 
extended to all $n$, would give the full evolution operator for any given 
function of the projection operators $|n\rangle \langle n|$. 
However, the initial semiclassical approximation is to consider the action-angle 
variables $(I,\theta)$ as appropriate conjugate variables for quantization.
From the action representation of the operator $U_N^l$, the $l^{th}$ power of
$U_N$ in \ (\ref{un}), the matrix elements in the angle representation turn
out to be
\begin{equation}
\label{me}
\langle \theta |U_N^l |\theta^\prime\rangle = {1 \over
2 \pi} \sum_{n=N_-}^{N_+}
\exp\left\{i\left[\left(n+{1\over 2}\right)(\theta-\theta^\prime)-l{\tau
E_n\over
\hbar}\right]\right\} \ ,
\end{equation}
where we used $\langle\theta | n\rangle = e^{-i\left(n+{1\over 2}\right)
\theta}/\sqrt{2\pi}$.  

The Poisson transformation is now applied, changing to the continuous action variable 
$I$, so that $E$ is interpolated by $H(I)$ as in \ (\ref{bs}).  
This results in 
\begin{equation}
\label{meI}
\langle \theta | U_N^l | \theta^\prime\rangle = {1\over 2\pi \hbar}
\sum_{m={-\infty}}^\infty (-1)^m \int_{{\cal I}_-}^{{\cal I}_+} 
{\rm d}I \exp\left\{{i\over \hbar} \Big[ I (\theta-\theta^\prime) + 
2\pi m - l\tau H(I) \Big] \right\} \ ,
\end{equation}
which is exactly equivalent to (\ref{me}).  Tracing over the angle
variables gives
\begin{equation}
\label{trace}
\mbox{Tr}(U_N^l) = {1\over \hbar} \sum_{m={-\infty}}^\infty (-1)^m 
\int_{{\cal I}_-}^{{\cal I}_+} {\rm d}I \exp\left\{{i\over \hbar} \Big[ 
2\pi mI - l\tau H(I) \Big] \right\} \ .
\end{equation}
The exponential in the integrand of (\ref{trace}) oscillates rapidly
in the semiclassical limit, and the largest contributions to the trace,
of order $\hbar^{1/2}$, come from the regions near the stationary phase
points at 
\begin{equation}
\label{points}
l\tau {{\rm d}H(I)\over {\rm d} I} -2 \pi m = 0 \ ,
\end{equation}
which is the same condition as the one given by (\ref{deriv}) for the 
periodic orbits of the classical stroboscopic map.  The stationary phase 
evaluation of the trace is therefore tied to the periodic orbits with actions $I_{m,l}$.

The periodic orbits can now be ordered by increasing period, $l$.
Equation (\ref{points}) also sets the range of possible repetitions
$(M_{l-}, M_{l+})$ of the period $l$ orbit. The frequencies, 
$\omega={{\rm d}H(I)/ {\rm d} I}$ in
the limited range of action $({\cal I}_-,{\cal I}_+)$ are also bounded. 
Thus, the number of periodic orbits in this range increases linearly with
$l$.  The full stationary phase evaluation of (\ref{trace}) becomes
\begin{equation}
\label{eval}
\mbox{Tr}(U_N^l) = \sum_{m=M_{l-}}^{M_{l+}} 
(-1)^m \left({2\pi  \over \hbar l \tau \left|{{\rm
d^2}H(I)\over {\rm d} I^2}\right|_{I_{m,l}}} \right)^{1/2}\!\!\!\exp\left\{
{i\over \hbar} \left[  2\pi mI_{m,l} - l\tau H(I_{m,l}) - \hbar{\pi\over
4}\right] \right\} \ .
\end{equation}
Not only are the stationary points of the integration specified by the
action variables of the periodic orbits, but the phase of each
contribution is given by the full action in units of $\hbar$,
\begin{equation}
\label{action}
{\cal S}_{m,l} = \int_0^{l\tau} {\rm d}t\ \left[ p \dot q - H\right ],
\end{equation}
except for the geometric ``Maslov term'', $\pi/4$.  

Typically, the contributions to the trace from the endpoints ${\cal
I}_\pm$ to the integrals in (\ref{trace}) are only of order $\hbar$. 
However, note that they cannot be separated from the stationary phase term
corresponding to a periodic orbit which lies very close to the boundary. 
In fact, by varying the parameters of the system and $l$, essentially
all the stationary phase points pass close to the boundaries.  The edge
corrections can be obtained as a uniform approximation~\cite{Eckhardt01}, but
this erases the simplicity of the periodic orbit expression for the
trace.  It is shown in the following section that the limitation of the
allowed phase space to the ring $({\cal I}_-,{\cal I}_+)$ in no way
hinders the Berry-Tabor equivalence.

%%%%%%%%%%%%%%%%%%%%%%%%%%%%%%%%%%%%%%%%%%%%%%%%%%%%%%%%%%%%%%%%%%%%%%%%%%%%%%%%%%%%%%%%%
\section{The Berry-Tabor equivalence}
\label{equiv}

A version of the Gutzwiller trace formula that is appropriate to an integrable 
map is simply obtained by Fourier analyzing the traces of $U_N^l$ in the discrete 
time $l$:
\begin{equation}
\label{bt}
\sum_{l=0}^\infty e^{il\theta} \mbox{Tr}(U_N^l) = \mbox{Tr}{1\over 1-
e^{i\theta}U_N} = \sum_n {1 \over 1- \exp [i(\theta-l\tau E_n)]} \ .
\end{equation}
Here, the eigenvalues of the map are considered as poles of the
resolvent $(1-zU_N)^{-1}$ instead of zeros of $\det(1-zU_N)$.

Though the infinite series of traces in (\ref{bt}) appears to be a
cumbersome alternative to evaluating the finite determinant, Poisson
transformation to the periodic orbit expression can be performed for
the semiclassical resolvent,
\begin{eqnarray}
\label{transform}
\mbox{Tr}{1\over 1-e^{i\theta}U_N} &=& \sum_{m,l} (-1)^m \left({2\pi \over\hbar
l \tau \left| {{\rm d^2}H(I)\over {\rm d} I^2}\right|_{I_{m,l}}}
\right)^{1/2} \nonumber\\
&& \times\exp \left( {i\over\hbar}\left[ \hbar l \theta + 2\pi m
I_{m,l} - l \tau H(I_{m,l}) -\hbar {\pi\over 4}\right]\right) \ .
\end{eqnarray}
Interpolating the discrete time $l$ again by the continuous time
$s$ gives
\begin{eqnarray}
\label{back}
\mbox{Tr}{1\over 1-e^{i\theta}U_N} &=& \sum_k (-1)^m \left({2\pi \over\hbar \tau l
\left| {{\rm d^2}H(I)\over {\rm d} I^2}\right|_{I_{m,s}} }\right)^{1/2}  
\nonumber\\
&& 
\times\int_0^\infty {\rm d}s\ \exp\left\{{i\over\hbar}\left[ \hbar s \theta +
2\pi (m I_{m,s} + \hbar ks) - s\tau H(I_{m,s}) -\hbar {\pi\over
4}\right]\right\} \ .
\end{eqnarray}
The stationary phase points, $s_{m,k}(\theta)$, of the integrals in
Eq.~(\ref{back}) are given by 
\begin{equation}
\label{statph2}
{{\rm d} {\cal S}_{m,s}\over {\rm d} s} + \hbar(\theta + 2 \pi k) = 0
\end{equation}
where ${\cal S}_{m,s}$ is the full action of the orbit singled out by
condition (\ref{points}).  It is important to note that, while in (\ref{points}),
the stationary phase condition defined a periodic orbit of integer period
$l$ for the $\tau$-stroboscopic map,  now $l$ has been replaced by the
continuous variable $s$.  This condition can be reinterpreted in two
ways: (i) orbits are picked that are no longer closed, or (ii) the
selected orbit is still periodic, but for a different
$\tau(\theta)$.  In either case, 
\begin{equation}
\label{condit}
{{\rm d} {\cal S}_{m,s}\over {\rm d} s} = \tau {{\rm d} {\cal
S}_{m,s}\over {\rm d} t} = \tau E_{m,s} \ ,
\end{equation}
where the energy of the selected orbit appears in the last equality. 
Thus, the resulting phase of each integral is given by the
reduced action for the orbits of energy $\hbar(\theta +2\pi k)/\tau$.  The
stationary evaluation of the resolvent is 
\begin{equation}
\label{resolv}
\mbox{Tr}{1\over 1-e^{i\theta}U_N} \approx \sum_{k,m} s_{1,k}(\theta) \>\exp \left[
{i\over \hbar} 2\pi mI\left({\hbar\over \tau}(\theta+2\pi k)\right) + i\pi
m\right] \ ,
\end{equation}
where $I(E)$ is the local inverse of $H(I)$.  Thus, if $\hbar/\tau$ is
chosen such that the variation of $\tau H(I)/\hbar$ does not exceed $2\pi$
for $I$ in $({\cal I}_-,{\cal I}_+)$, there are at most a few orbits
contributing to the semiclassical resolvent for each value of
$\theta$.  Since the repetitions must still be summed, the interpretation
(ii) for (\ref{statph2}) as a periodic orbit of another stroboscopic map
is perhaps the more appealing.  

The condition for the resolvent to be singular is that all the $m$
repetitions are exactly in phase,
\begin{equation}
\label{bs2}
I\!\left({\hbar\over \tau}(\theta +  2\pi k)\right) = \left(j-{1\over 2}\right)\pi \ ,
\end{equation}
which is just the initial Bohr-Sommerfeld quantization.  Thus, the
periodic orbit evaluation of the map resolvent retrieves the full
Berry-Tabor equivalence without any need to consider boundary corrections
to the trace.  The reason why this works is that the singularities of the
resolvent only manifest themselves by summing the periodic contributions
for very long times, or multiple repetitions.  For these contributions
there is an effective increase of the large parameter of the stationary
phase evaluation of $\mbox{Tr}(U_N^l)$ given by (\ref{eval}).  The integrals
are then dominated by a very narrow region near each periodic orbit,
so that the boundary can be ignored for the high repetition of an orbit
even if it lies almost at the edge.

The evaluation of the spectral determinant may be regarded as a
resummation of the semiclassical sum for the poles of the resolvent, such
that the (discrete) time  of the contributing periodic orbits is cut-off
by the dimension of the finite Hilbert space.  We show in the next
section that the error in the traces of the propagator due to the edge corrections
can be magnified by the spectral determinant.

%%%%%%%%%%%%%%%%%%%%%%%%%%%%%%%%%%%%%%%%%%%%%%%%%%%%%%%%%%%%%%%%%%%%%%%%%%%%%%%%%%%%%%%%%%%%%%%%%%%
\section{Model Hamiltonian}
\label{model}

To illustrate the calculation of the spectral determinant, we choose the
simplest Hamiltonian that is a nonlinear function of the action, namely
\begin{equation}
\label{ham}
H(I) = {1\over 2} (I-{\cal I})^2 \qquad (I\ge 0) \ .
\end{equation}
The periodic orbit condition (\ref{deriv}) then reduces to 
\begin{equation}
\label{simple}
I_{m,l} = {\cal I} + 2\pi {m\over l  \tau} \ 
\end{equation}
and the stationary phase evaluation of the trace assumes the explicit form
\begin{equation}
\label{spsimp}
\mbox{Tr}(U_N^l) \approx \left({2 \pi \over \hbar l\tau}\right)^{1/2}
e^{-i\pi/4} \sum_m (-1)^m \exp \left[{i\over \hbar} 2\pi m\left( {\cal I}
+ {\pi m \over l \tau}\right)\right] \ .
\end{equation}

For the case where the parameter ${\cal I} = \hbar (n+1/2)$ (i.e. one of
the quantized actions), $\mbox{Tr}(U_N^l)$ reduces to a ``curlicue''.  These
recursively spiraling patterns in the complex plane were shown by Berry
and Goldberg~\cite{Berry88} to have quite diverse characteristics depending
on $\tau/\hbar$, in the limit $N\rightarrow \infty$.  This may also be
the case for non-quantized ${\cal I}$.

Two essentially different regions of the eigenspectrum can be studied.  
In case (i), ${\cal I}>0$ and ${\cal I}_\pm={\cal I}\pm \sqrt{2\cal E} < 2
{\cal I}$.  The energy levels there originate on two different branches of
$H(I)$, so there may be near degeneracies.  This is a similar situation to
that expected for an integrable system with more degrees of freedom
(see e.g.~\cite{text}).
Otherwise, in case (ii), both ${\cal I}_+$ and ${\cal I}_-$ $>$ $2{\cal I}$, 
so that a single branch of the $H(I)$ curve is sampled.  The energy levels 
are then quite regularly spaced with approximate separation of
Planck's constant times the average ${\rm d}H/{\rm d}I$ for this energy interval.
By choosing $E_\pm$ to satisfy condition (\ref{constraint}),
we also guarantee the regularity of the spectrum of $U_N$ itself.

%%%%%%%%%%%%%%%%%%%%%%%%%%%%%%%%%%%%%%%%%%%%%%%%%%%%%%%%%%%%%%%%%%%%%%%%%%%%%%%%%%%%%
\subsection{Case (i)}

\begin{figure}
\includegraphics[width=11.0cm]{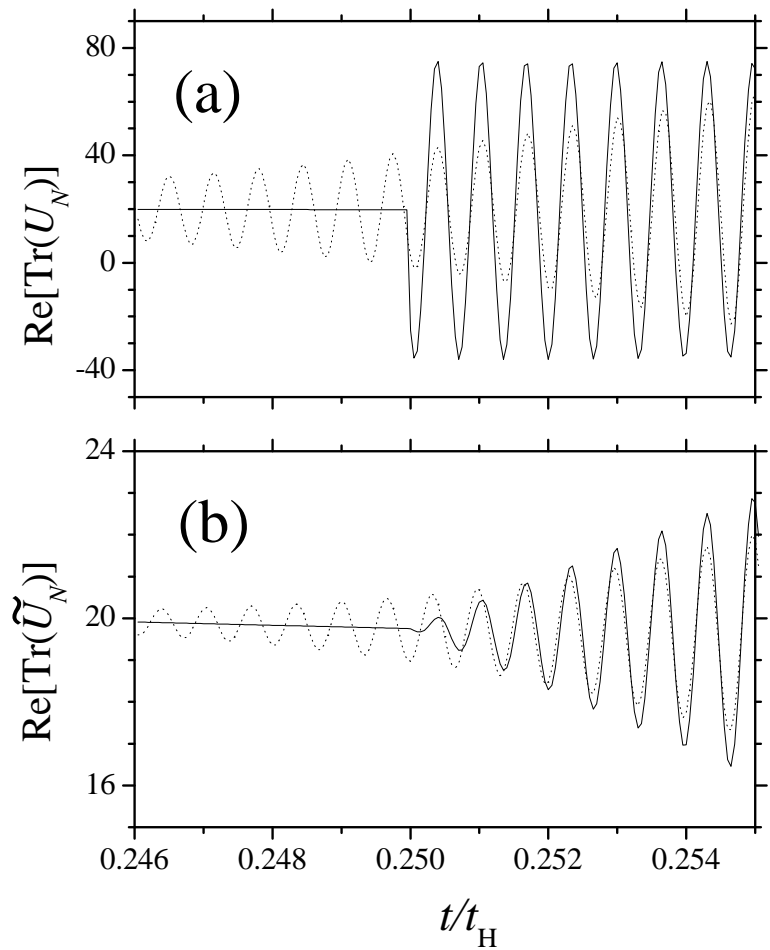}
\caption{Comparison between the exact (dotted lines) and the semiclassical values 
(solid lines) of Re[Tr($U_N)$] as a function of $t/t_{\mbox{\scriptsize H}}$ for 
(a) unweighted and (b) cosine weighted trace. 
The behaviour is similar for Im[Tr($U_N)$] and therefore not shown here.}
\label{one}
\end{figure}

Let us first examine the semiclassical approximations to $\mbox{Tr}(U_N)$ for a single
period. Notice that the stationary phase approximation to the integrals in
(\ref{trace}) would be exact for the quadratic model if the limits
could be extended to $\pm \infty$.  In other words, the edge corrections
are the only deviation of the periodic orbit approximation with respect to
the exact trace.  These corrections are certainly not negligible for stroboscopic 
parameters such that a periodic orbit is close to one of the edges. 
The simplest asymptotic edge corrections, as deduced in~\cite{Eckhardt01}, are presented 
in \ref{sec:appA}. The computed error due to 
the edges is displayed in   
Figure \ref{one}. Here we let the time $t$ vary 
and study Tr($U_N$) as a function of $t/t_{\mbox{\scriptsize H}}$, where 
$t_{\mbox{\scriptsize H}}=2\pi\Delta N/E$ is the Heisenberg time. 
As the time is increased the integrand at the l.h.s.~of Equation (\ref{trace}) has more
stationary phase points.
The sudden jump in the semiclassical trace in Figure \ref{one}a is due to the following.
For small values of $t/t_{\mbox{\scriptsize H}}$ there is just a single stationary phase 
point with $m=0$. Going beyond $t/t_{\mbox{\scriptsize H}} \approx 0.25$ 
all $|m|\le 1$ become stationary phase points. (It is near such transition points that the
semiclassical approximation is worse.) This situation repeats itself regularly
with increasing $t/t_{\mbox{\scriptsize H}}$ adding more stationary points to 
the sum (\ref{spsimp}).
As seen in Figure~\ref{one}a, the semiclassical trace as given by (\ref{spsimp}) is not 
an accurate approximation to the exact Tr$(U_N)$.
The suppression of the edge correction by means of a cosine weighted trace,
namely
\begin{equation}
\mbox{Tr}(\widetilde{U_N}) = \mbox{Tr} \left[U_N \cos\left(\frac{\pi 
H(I)}{2E}\right)\right],
\end{equation}
improves the agreement enormously, as shown in Figure \ref{one}b.
Our calculations have been performed for case $\hbar= 0.01, E = 30.4571694,$
and ${\cal I} = 20.3489573$ giving $N=1~521$.

\begin{figure}
\includegraphics[width=11.0cm]{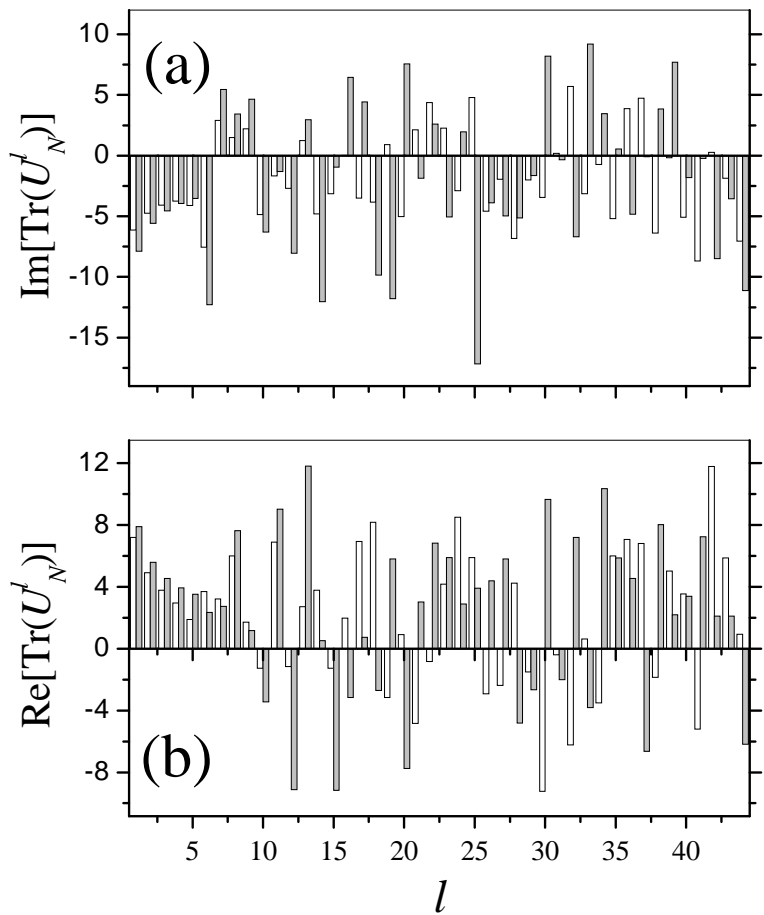}
\caption{Comparison between the exact (white histograms) and the semiclassical 
(grey histograms) values of (a) Im[Tr$(U_N^l$)] and (b) Re[Tr$(U_N^l$)] as a 
function of $l$ for $N=48$.}
\label{fig:tracesN=48}
\end{figure}

We now return to the original programme of obtaining the eigenvalues of $U_N$,
defined for a fixed $\tau$, by resumming periodic orbits.
The first step is to calculate the traces of $U_N^l$. The coefficient of the 
characteristic polynomial are then easily computed by (\ref{eq:recurrence}).
The roots of the resulting polynomial are obtained numerically. 
The precision in finding the roots, limits this method to polynomials 
with $N \le 50$.
We chose the model Hamiltonian parameters accordingly.

To illustrate case (i), we take the parameters $E$ and ${\cal I}$ to be the same
as above and $\hbar = 0.35$, which gives $N=48$. 
Figure~\ref{fig:tracesN=48} displays the real and imaginary parts of 
$\mbox{Tr}(U_N^l)$ for $l\le N$ using the semiclassical expression 
(\ref{spsimp}) contrasted with their exact values. 
As expected, the semiclassical approximation works reasonably well for the lowest 
values of $l$ and gives very poor results for the largest ones.
It is worth noticing that similar calculations with the addition of an imaginary 
part to the time in the trace (\ref{eval}) show a dramatic improvement in the 
agreement between the semiclassical and the exact traces.
This is further evidence that the differences are due to edge corrections
and not to numerical inaccuracies.

\begin{figure}
\includegraphics[width=11.0cm]{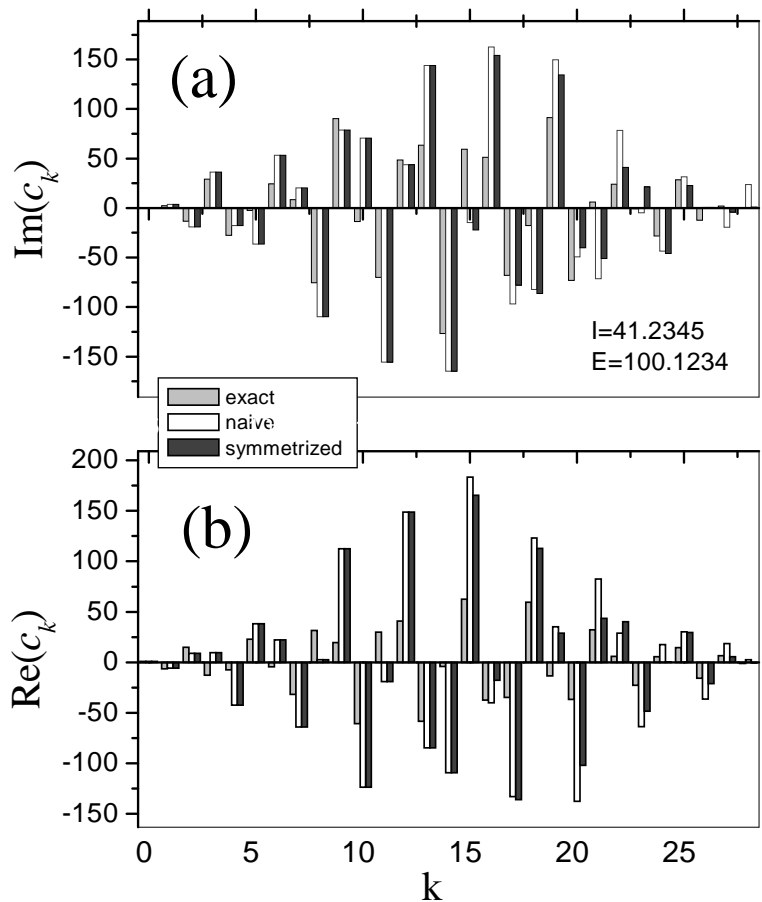}
\caption{Comparison between the exact (gray), the semiclassical (white) and the 
symmetrized semiclassical (black) values of (a) Im($c_k$)  and (b) Re($c_k$) as a 
function of $k$ for $N=28$.}
\label{fig:coefN=28}
\end{figure}

Next, we calculate the coefficients $c_k$ of the characteristic polynomial
(\ref{char}) using the recursion relation (\ref{eq:recurrence}).
The symmetry relation (\ref{symm}) allows us to use only traces corresponding
to times shorter than $\tau_{\mbox{\scriptsize H}}/2$.
In Figure \ref{fig:coefN=28} we compare the exact coefficients $c_k$ with
those resulting from the semiclassical traces, with and without
symmetrization. To better illustrate the whole range of $c_k$ we show a 
situation where $N=28$, corresponding to $\hbar= 1.0, E = 100.1234,$
and ${\cal I} = 41.2345$.
As before, the semiclassical approximation works well for the low $k$ coefficients
($k < 5$) and, obviously, also for the corresponding higher $k$'s if we impose the  
symmetry (\ref{symm}). 
For the remaining coefficients, the agreement with the exact $c_k$ is poor.

\begin{figure}
\includegraphics[width=11.0cm]{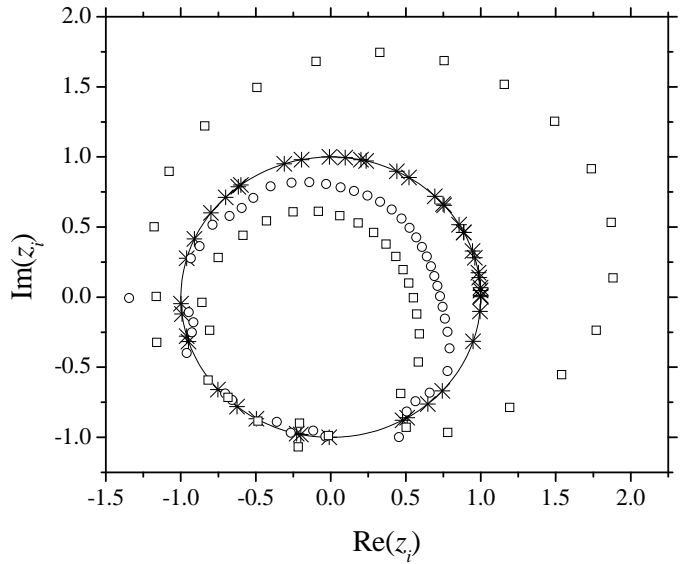}
\caption{Comparison between the exact (asterisks), the semiclassical (open circles) 
and the symmetrized semiclassical (open squares) roots $z_i$ of the characteristic 
polynomial for $N=48$.}
\label{fig:rootsN=48}
\end{figure}

Finally we arrive at the eigenvalues of $U_N^l$.
Figure \ref{fig:rootsN=48} shows the results for $\hbar= 0.35, E = 30.4571694,$
and ${\cal I} = 20.3489573$, which is a rather typical situation. 
The exact roots of the characteristic polynomial lie on the unit circle, as they
should. The standard semiclassical approximation destroys unitarity and the roots of 
the corresponding characteristic polynomial are no longer restricted to the unit circle.
The enforcement of the symmetry (\ref{symm}) makes $|\det(U)| = 1$.
As a consequence the roots either lie exactly on the unit circle, or appear in pairs, 
one inside and one outside the circle. 
More precisely, the self-inversive symmetry of the characteristic polynomial renders
a symmetry in its zeros: if $z_k$ is a root then $1/z^*_k$ is also a root.
Indeed, it has been shown by Bogomolny et al.\cite{BBL96} that on average only about
$57\%$ of the roots of a random symmetrical polynomial lie on the unit circle.
Unfortunately, upon symmetrization individual roots do not necessarily come closer 
to the exact ones, as compared with the standard procedure. Figure 4 shows that they can
even be pushed out, unless the standard semiclassical roots are
already close to the exact ones, in which case symmetrization tends to improve the
accuracy. Unfortunately this behaviour is not very systematic.

%%%%%%%%%%%%%%%%%%%%%%%%%%%%%%%%%%%%%%%%%%%%%%%%%%%%%%%%%%%%%%%%%%%%%%%%%%%%%%%%%%%%%
\subsection{Case (ii)}

\begin{figure}
\includegraphics[width=13.0cm]{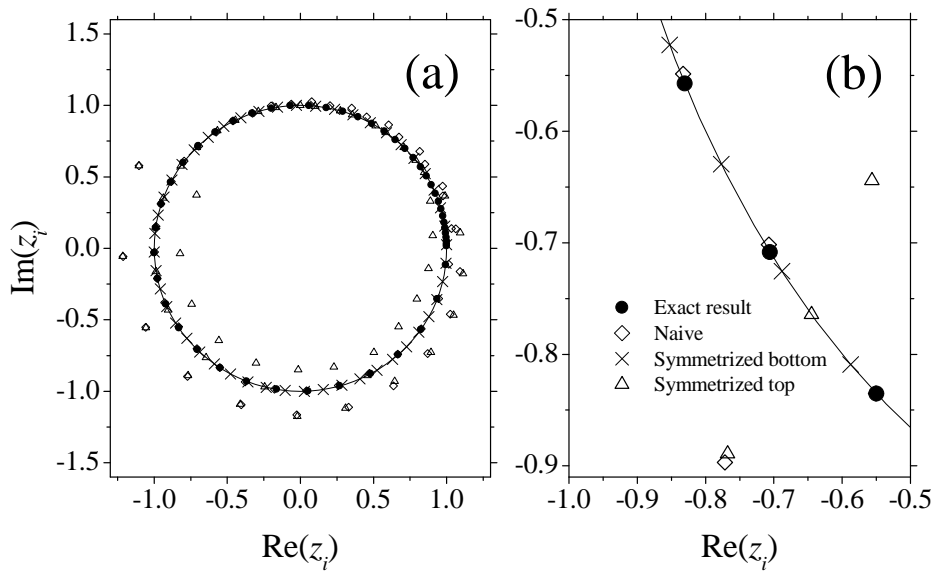}
\caption{Comparison between the exact (crosses), the semiclassical (open circles), 
the ``bottom" symmetrized semiclassical (open squares) and the ``top" symmetrized 
semiclassical (open triangles) roots $z_i$ of the characteristic polynomial for 
$N=48$. Panel (a) displays all roots, while (b) is a scale blow up of the same
data showing that the nice agreement is only apparent.}
\label{fig:rootscusp}
\end{figure}

We switch now to case (ii) and consider the situation where the energy 
$E>{\cal I}^2/2$, or equivalently, both ${\cal I}_+$ and ${\cal I}_->2{\cal I}$.
As mentioned before, in this situation the levels are almost equally spaced with
separation of $\hbar {\rm d}H/{\rm d}I$. 
From the semiclassical point of view, the most important distinction to case (i) 
is the absence of the $m=0$ orbit.
Actually, for our simple model there will be no tori with orbits of period $l<N/2$
in case (ii). Indeed, the fraction, $f$, of zero traces in the semiclassical
approximation is easily seen to be: 
\begin{equation}
f=\frac{1}{2}\frac{\sqrt{E_+}+\sqrt{E_-}}{\sqrt{E_+}-\sqrt{E_-}}.
\label{fraction}
\end{equation}
The optimum of $f\approx N/2$ is approached for $E_-\approx 0$, which implies 
${\cal I}\leq 0$,
so we choose $\hbar= 1.0, {\cal I} = -1.03489573$ with $E_ -= 0.265678$
and $E_+=1250.0$, which corresponds to $N=48$.
Keeping $\tau$ as defined by (\ref{constraint}) gives Tr$(U_N^l)=0$ for 
$l \le 24$. 
The results are summarized by Figure \ref{fig:rootscusp}.

As in case (i) the simple semiclassical approximation shows roots outside
the unit circle. The standard semiclassical improvement by symmetrization 
(``bottom up symmetrization")
erases all system information, since there are only nonzero traces for long
times. Indeed, the spectral determinant (\ref{expan}) reduces to
\begin{equation}
P_N (z)=1+(-1)^N \det(U_N)\>z_N.
\label{expan1}
\end{equation}
Instead, one can maximize the semiclassical information by symmetrizing the 
lower coefficients in $P_N(z)$ from the higher $c_k$'s , containing the 
traces with long orbits (``top down symmetrization"). However, the final 
result is no better than the one without symmetrizing at all. 

By shifting the parameters in case (ii) so that $H(I)$ becomes almost linear, 
within the interval $({\cal I}_-,{\cal I}_+)$, the  agreement
between the different approaches becomes much more reasonable 
than in Figure \ref{fig:rootscusp}.
However, if one recalls that the mean level spacing is known and the levels 
are almost equally spaced, some caution is then required.
The exact traces lead to a characteristic polynomial with self-inversive symmetry.
For small values of $l$, where there is no stationary phase and the semiclassical 
traces are zero, the (modulus of) exact traces are small compared to unit.
On the other hand, we observe that, for large $l$ values, the semiclassical approximation 
fails to reproduce the exact traces with the same precision.
Thus, the best semiclassical agreement corresponds surprisingly to the case where all 
traces are taken as zero (bottom up). 
As shown in Fig \ref{fig:rootscusp}, the top down symmetrization is superior to the 
standard procedure only where the eigenvalues are sufficiently accurate.
It then brings most of the roots back to the unit circle. However, since 
it employs some inaccurate large $l$ traces, (top down) symmetrization also produces 
pairs of roots lying
at opposite sides of the unit circle, like in case (i).

%%%%%%%%%%%%%%%%%%%%%%%%%%%%%%%%%%%%%%%%%%%%%%%%%%%%%%%%%%%%%%%%%%%%%%%%%%%%%%%%%%%%%%%
\section{Conclusions}

The main assets of using spectral determinants to resum the periodic orbits 
in the Gutzwiller trace formula for chaotic quantum maps are: 
(a) the requirement of unitarity is partially incorporated, 
(b) the necessity of handling very long (and exponentially many) periodic orbits 
    with periods larger than $\tau_H/2$ is eliminated and 
(c) the method produces isolated levels rather than a smoothed density of states.
Nevertheless, the method remains problematic in general, because of the necessity 
of accounting for a formidable number of periodic orbits in the semiclassical 
limit, which is a daunting task from the classical point of view.

Even though semiclassical spectra of integrable systems are obtained most efficiently
by generalizations of the Bohr-Sommerfeld rules, one could 
expect that spectral determinants would still be more successful
tools for integrable spectra than the trace formula, since they take some account of 
unitarity and here the classical periodic orbit structure is much 
simpler than for chaotic systems.
Our study shows that this is not the case.

We find large errors in the semiclassical traces, which could, in principle,
be completely fixed by accounting for edge corrections. Indeed, this has been verified
in the case of the first order corrections in the Appendix. However, to follow such a path
would be at odds with the spirit of the present work. Rather than solving a trivial model,
our purpose was to assess the efficacy of the spectral determinant, symmetrized or not,
as a tool of reducing the effect of inaccuracies of the semiclassical
traces in the calculation of eigenvalues. The spectrum of our model is exactly known
and it allows complete control over each stage of the approximation.
Thus, we have worked solely within a framework that is based on the usual amplitudes
of periodic orbits in the trace formula. Even though our map has been obtained
by looking stroboscopically at a simple system with continuous time and its boundaries 
are entirely arbitrary, it is arguable that our results may resemble those for a quantum map
that results by taking a ``Bogomolny section" of a Hamiltonian system 
\cite{Bogomolny90, Bogomolny92}. The corresponding classical map must also have a boundary
and it will also be integrable,
if this is a property of the original Hamiltonian system.

The errors in the semiclassical traces, discussed in Section \ref{model},
yield inaccurate coefficients for the characteristic polynomial, 
rendering poor approximations for the spectra of integrable systems.
The symmetry property of the spectral determinant does not necessarily 
lead to better approximations. Actually in most of the cases studied 
there is only improvement where the unsymmetrized eigenvalues were already 
reasonably accurate. We can now understand this to account for the encouraging results
obtained by Creagh \cite{Creagh95} for the perturbed cat map. Even though the exact eigenvalues
are not determined semiclassically as in our case, his system lies very close to the
linear map, where the periodic orbit traces are exact, so that the symmetrization always
fixes the eigenvalues on the unit circle. In contrast, if one perturbs the traces in the present
integrable system continuously from their exact values to their semiclassical approximation, 
the eigenvalues may collide on the unit circle and be knocked off as a pair, 
of which we see many examples in our model. We have also tried fitting the spectrum by
treating $\det U_N$ as a free parameter, without any essential qualitative improvement
of the results.

It is important to mention that the edge correction may be neglected 
if the eigenvalues are obtained  by Fourier transformation over long 
times, so that the Berry-Tabor equivalence does not need to be ``dressed" 
because of the boundary. Of course, this is not much help if the system 
is not integrable. Indeed, an important point concerning the Bogomolny 
approach is that no assumption is made about characteristics of the
dynamics. Recently, this method has also been extended successfully to 
describe the eigenstates themselves of a chaotic system \cite{SS00}. 
The present study of the integrable limit, though not obtained by a 
surface of section, suggests that caution may be required in any attempt 
to extrapolate the general section-method to nearly integrable,
or mixed systems.

\section*{Acknowledgments}
%\ack

We gratefully acknowledge discussions with M.~Saraceno.  AMOA and CHL
acknowledge support from CNPq, and ST acknowledges support by the US
National Science Foundation under Grant No. PHY-0098027, and the
hospitality of CBPF and UERJ, Brazil.  

%%%%%%%%%%%%%%%%%%%%%%%%%%%%%%%%%%%%%%%%%%%%%%%%%%%%%%%%%%%%%%%%%%%%%%%%%%%%%%%%%%%%%%%
\appendix
\section{Edge corrections}
\label{sec:appA}

The asymptotic form for the edge corrections follows by basic application
of the expansion
\begin{equation}
\label{edgeasym}
C(z) + i S(z) = \int_0^z {\rm d}t e^{i{\pi \over 2} t^2}  \sim {e^{i{\pi
\over 4}} \over \sqrt{2}} {\rm sign}(z) - {i e^{i{\pi \over 2} z^2} \over
\pi z} + {\rm O}\left( |z|^{-3} \right) \ .
\end{equation}
where the first term gives the stationary phase contribution; actually,
it is the complex conjugate with $z$ real which turns out to be needed. 
The second term is responsible for the edge corrections.  Expanding the
argument of the exponential in (\ref{trace}) to quadratic order in
$I^\prime = I - I_{m,l}$ for the Hamiltonian in (\ref{ham}), gives
\begin{equation}
\label{argument}
Arg\ = {i \over \hbar}\left( 2\pi m \varphi  + {2\pi^2 m^2 \over 
 l \tau }  - { l\tau \over 2 } {I^\prime}^2\right)
\end{equation}
Rescaling the action variable by $\sqrt{\pi\hbar /l\tau}$ matches the
argument of (\ref{edgeasym}) with that of (\ref{argument}).  After a
little algebra, the edge corrections $\epsilon$ take the form
\begin{equation}
\label{corre}
\epsilon \sim {i\over l \tau} \sum_m \exp\!\!\left\{{i\over \hbar} \left(2\pi m
\varphi + {2 \pi^2 m^2 \over l\tau} \right)\right\} \left[ {1\over {\cal I}_+}
\exp\!\!\left(-i{l\tau \over 2\hbar} {\cal I}_+^2\right) - {1\over {\cal I}_-} 
\exp\left(-i{l\tau \over 2\hbar} {\cal I}_-^2\right) \right]
\end{equation}
For a discussion of edge corrections for the general case where the phase
in the integral (\ref{trace}) is not quadratic, see~\cite{Eckhardt01}.


\begin{thebibliography}{99}

\bibitem{cv} Cvitanovic P, Percival I and Wirzba A 1992 (editors), Chaos
focus issue on periodic orbit theory, {\it Chaos} {\bf 2}, (1992).

\bibitem{Cvitanovic89} 
   Cvitanovic P and Eckhardt B 1989 
        {\it Phys.~Rev.~Lett.} {\bf 63} 823

\bibitem{Eckhardt89} 
   Eckhardt B and Aurell E 1989 
        {\it Europhys.~Lett.} {\bf 9} 509

\bibitem{Gutzwiller71} 
   Gutzwiller M C 1971 {\it J.~Math.~Phys.} {\bf 12} 343

\bibitem{Balian72} 
   Balian R and Bloch C 1972 
        {\it Ann.~Phys.~(N.Y.)} {\bf 69} 76 

\bibitem{Berry90JPA} 
   Berry M V and Keating J P 1990 
        {\it J. Phys. A: Math. Gen.} {\bf 23} 4839
	
\bibitem{Bogomolny90} 
   Bogomolny E B 1990 {\it Comments At.~Mol.~Phys.}~{\bf 25} 67

\bibitem{Bogomolny92} 
   Bogomolny E B 1992 {\it Nonlinearity} {\bf 5} 805

\bibitem{Eckhardt01} 
   Eckhardt B and Smilansky U 2001 
        {\it Found.~Phys.} {\bf 31} 543

\bibitem{bt} 
   Berry M V and Tabor M 1976 
        {\it P.~Roy.~Soc.~Lond.~A Mat.} ~{\bf 349} 101; 1977 
        {\it J. Phys. A: Math. Gen.} {\bf 10} 371

\bibitem{text} 
   Ozorio de Almeida A M, {\it Hamiltonian Systems: Chaos
and Quantization}, (Cambridge University Press, Cambridge, 1988).

\bibitem{Creagh95} 
   Creagh S C 1995, 
        {\it Chaos} {\bf 5} 477

\bibitem{Saracero92} 
   Saraceno M and Voros A 1992
        {\it Chaos} {\bf 2} 99

\bibitem{Smilansky93JPA} 
   Smilansky U 1993 {\it Nucl. Phys.} A {\bf 560} 57

\bibitem{Berry85} 
   Berry M V 1985 
        {\it Proc. R. Soc. Lond.} A {\bf 400} 229

\bibitem{Berry88} 
   Berry M V and Goldberg J 1988 
        {\it Nonlinearity} {\bf 1} 1

\bibitem{BBL96} 
   Bogomolny E, Bohigas O and Leboeuf P 1996 
        {\it J. Stat. Phys.} {\bf 85} 639	 
\bibitem{SS00}
   Simonotti F P and Saraceno M 2000
        {\it Phys. Rev. E} {\bf 61} 6527 

\end{thebibliography}
\end{document}